\documentstyle[preprint,tighten,aps]{revtex}
\begin{document}
\draft
\baselineskip = 0.87\baselineskip
\title
{Universal critical coupling constants \\ for the three--dimensional  
$n$--vector model \\ from field theory} 

\author{A. I. Sokolov$^{1,2}$, ~E. V. Orlov$^1$,  
~V. A. Ul'kov$^2$, \\ and ~S. S. Kashtanov$^1$}

\address
{$^1$Department of Physical Electronics, Saint Petersburg 
Electrotechnical University, \\
Professor Popov Street 5, St. Petersburg, 197376, Russia \\
$^2$Department of Physics, Saint Petersburg Electrotechnical University, \\
Professor Popov Street 5, St. Petersburg, 197376, Russia}

\maketitle

\begin{abstract}
The field-theoretical renormalization group (RG) approach in three dimensions
is used to estimate the universal critical values of renormalized coupling 
constants $g_6$ and $g_8$ for the $O(n)$-symmetric model. The RG series 
for $g_6$ and $g_8$ are calculated in the four-loop and three-loop 
approximations, respectively, and then resummed by means of the 
Pad\'e-Borel-Leroy technique. Under the optimal value of the shift 
parameter $b$ providing the fastest convergence of the iteration procedure, 
numerical estimates for $g_6^*$ are obtained with the accuracy no worse 
than 0.3$\%$. The RG expansion for $g_8$ demonstrates a stronger 
divergence, and results in considerably cruder numerical estimates.

\vspace{1.5cm}

PACS numbers: 05.70.Jk, 11.10.Gh, 64.60.Ak, 64.60.Fr
 
\end{abstract}

\newpage

\section{Introduction}
\label{sec:1}

The three-dimensional (3D) $O(n)$-symmetric model plays a very important 
role in the theory of phase transitions. It describes critical phenomena in 
a variety of physical systems including Ising, XY-like and Heisenberg 
ferromagnets, simple fluids and binary mixtures, superconductors and Bose 
superfluids, etc. This model is also relevant to certain asymptotic regimes 
of the critical behavior of the quark-gluon plasma in quantum chromodynamics 
($n = 4$) \cite{W,RW}. In the critical region, the $n$-vector model is known 
to be thermodynamically equivalent to the 3D Euclidean field theory of 
$\lambda \varphi^4$ type, and may be treated by the field-theoretical 
renormalization group (RG) technique which proved to be very efficient both 
for studying the qualitative features of phase transitions and calculating 
the critical exponents \cite{BNM,LGZ,ZJ,AS,S98}. 

On the other hand, for decades the influence of ordering fields upon 
the critical behavior of various systems attracted permanent attention, 
being of prime interest both for theorists and experimentalists. Recently, 
the free energy (effective action) and, in particular, higher-order 
renormalized coupling constants $g_{2k}$ for the basic models of 
phase transitions became the target of intensive theoretical studies 
\cite{S98,BB,B1,B2,TW,Ts94,B3,R,S96,ZLF,SUO,SOU,GZ,Morr,BC97,Ts97,PV}. 
These constants are related to the non-linear susceptibilities 
$\chi_{2k}$ and enter the scaling equation of state, thus playing a key 
role at criticality. Along with critical exponents and 
critical amplitude ratios, they are universal, i.e. they possess, under 
$T \to T_c$, numerical values that are not sensitive to the physical 
nature of the phase transition, depending only on the system dimensionality 
and the symmetry of the order parameter. 

Calculation of the universal critical values of $g_6$, $g_8$, etc. 
for the three-dimensional Ising model by a number of analytical and 
numerical methods showed that the field-theoretical RG approach in fixed 
dimensions yields the most accurate numerical estimates for these 
quantities. It is a consequence of a rapid convergence of the 
iteration schemes originating from renormalized perturbation 
theory. Indeed, the resummation of four- and five-loop RG 
expansions by means of the Borel-transformation-based procedures
gave the values for $g_6^*$, which differ from each other by less
than $0.5 \%$ \cite{SOU,GZ}, while the use of a resummed three-loop RG 
expansion enabled one to achieve an apparent accuracy no worse than 
$1.6 \%$ \cite{S98,SUO}. Moreover, the field-theoretical RG approach 
turns out to be powerful enough even in two dimensions: properly resummed 
four--loop RG expansions lead to fair numerical estimates for the critical 
exponents \cite{BNM} and the renormalized coupling constant $g_6^*$ \cite{SO} 
of a 2D Ising model, and give reasonable results for its random counterpart 
\cite{MSS,M}. It is natural, therefore, to use the field theory for 
a calculation of renormalized higher-order coupling constants for 
the 3D $n$-vector model. In this paper, the 3D RG expansion for the 
renormalized coupling constants $g_6$ and $g_8$ will be calculated, and the 
numerical estimates for their universal critical values will be obtained.

\section{RG expansions for the sextic and octic coupling constants}
\label{sec:2}

Within the field-theoretical language, the 3D  
O(n)-symmetric model in the critical region is described by Euclidean 
scalar field theory with the Hamiltonian
\begin{equation}
H = 
\int d^3x \Biggl[{1 \over 2}( m_0^2 \varphi_{\alpha}^2
 + (\nabla \varphi_{\alpha})^2) 
+ \lambda (\varphi_{\alpha}^2)^2 \Biggr] ,
\label{eq:1} \\
\end{equation}
where a bare mass squared $m_0^2$ is proportional to 
$T - T_c^{(0)}$, $T_c^{(0)}$ being the phase transition
temperature in the absence of the order parameter fluctuations. 
Taking fluctuations into account results in renormalizations
of the mass $m_0 \to m$, the field $\varphi \to \varphi_R$, 
and the coupling constant $\lambda \to m g_4$. Moreover,
thermal fluctuations give rise to many-point correlations 
$<\varphi(x_1) \varphi(x_2)...\varphi(x_{2k})>$ and, 
correspondingly, to higher-order terms in the expansion of the 
free energy in powers of the magnetization $M$:
\begin{equation}   
F(M, m) = F(0, m) + \sum_{k = 1}^{\infty} \Gamma_{2k} M^{2k}.
\label{eq:2} \\
\end{equation}
In the critical region, the coefficients $\Gamma_{2k}$, being 
one-particle irreducible $2k$-point vertices taken at zero external 
momenta, demonstrate well-known scaling behavior:  
\begin{equation}
\Gamma_{2k} = g_{2k} m^{3 - k(1 + \eta)},
\label{eq:3} \\
\end{equation}
where $\eta$ is a Fisher exponent, and $g_{2k}$ are some constants.
Let us set as usually $g_2 = 1/2$. Then $g_4$, $g_6$, $g_8$,... will 
acquire universal values. The asymptotic critical values of $g_4$, 
$g_4^*(n)$, determine the critical exponents and other universal
quantities, thus playing a very important role in the theory. The numbers 
$g_4^*(n)$ have been found by resummation of the six-loop expansion for the 
RG $\beta$-function \cite{BNM,LGZ,AS,S98}, from strong-coupling series  
\cite{CPRV}, by lattice calculations \cite{BC97}, from the 
$\epsilon$-expansion \cite{PV7}, and are known today with an 
accuracy which may be considered rather high.

The universal values of higher coupling constants $g_6^*$, $g_8^*$, etc. 
determine the structure of the free energy $F(M,m)$ under strong critical 
fluctuations. In fact, the Taylor expansion of the scaling
function contains the ratios $g_{2k}^*/(g_4^*)^{k-1}$, which may be 
easily shown by replacement of the magnetization $M$ in Eq.(\ref{eq:2}) 
by the dimensionless variable $z = M \sqrt{g_4/m^{1 + \eta}}$:
\begin{equation}
F(z,m) - F(0,m) = {m^3 \over g_4} \Biggl({z^2 \over 2} + z^4 
+ {g_6 \over {g_4^2}} z^6 + {g_8 \over {g_4^3}} z^8 + ... \Biggr).
\label{eq:4} \\
\end{equation}
Moreover, via $g_{2k}$, the non-linear susceptibilities $\chi_{2k}$
can be expressed. For $\chi_4$ and $\chi_6$ corresponding formulas
are as follows: 
\begin{equation}
\chi_4 = {\partial^3M \over{\partial H^3}} \Bigg\arrowvert_{H = 0} 
= - 24 \chi_2^2 m^{-3} g_4, \qquad \quad 
\chi_6 = {\partial^5M \over{\partial H^5}} \Bigg\arrowvert_{H = 0} 
= 720 \chi_2^3 m^{-6}(8 g_4^2 - g_6).
\label{eq:5} \\
\end{equation}
Their inversion gives the relations 
\begin{equation}
g_4 = - {{m^3 \chi_4} \over{24 \chi_2^2}}, \qquad \qquad \qquad
g_6 = {{m^6 (10 \chi_4^2 - \chi_6 \chi_2)} \over{720 \chi_2^4}},
\label{eq:6} \\
\end{equation}
which are widely used for extraction of numerical values of renormalized 
coupling constants from the results of lattice calculations 
\cite{R,ZLF,BC97,Ts97,B,BK}.

The method of calculating the RG series for the $g_6$ and $g_8$ we use here 
is straightforward. Since in three dimensions higher-order bare couplings are 
irrelevant in RG sense, the renormalized perturbative series to be found can 
be obtained from conventional Feynman graph expansions for the six-point and 
eight-point vertices in terms of the only bare coupling constant -- $\lambda
$. In the course of calculations the tensor structure of these vertices, 
\begin{equation}
\Gamma_{\alpha \beta \gamma \delta \mu \nu} = {1 \over 15} 
(\delta_{\alpha \beta} \delta_{\gamma \delta} \delta_{\mu \nu} +
14 ~transpositions) \Gamma_6.
\label{eq:7} \\
\end{equation}
\begin{equation}
\Gamma_{\alpha \beta \gamma \delta \mu \nu \rho \sigma} = {1 \over 105} 
(\delta_{\alpha \beta} \delta_{\gamma \delta} 
\delta_{\mu \nu} \delta_{\rho \sigma} +
104 ~transpositions) \Gamma_8.
\label{eq:8} \\
\end{equation}
should be taken into account. In its turn, $\lambda$ 
may be expressed perturbatively as a function of the renormalized 
coupling constant $g_4$. Substituting corresponding power series for 
$\lambda$ into original expansions, we can obtain the RG series for $g_6$ 
and $g_8$. The one-, two-,  
three- and four-loop contributions to $g_6$ are formed by one, three, 16, 
and 94 one-particle irreducible Feynman graphs, respectively. Their 
calculation gives:
\begin{eqnarray}
g_6 = {9 \over \pi}{\Biggl({\lambda Z^2 \over m} \Biggr)^3}
\Biggl[ {n + 26 \over {27}}  -  {9~n^2 + 340~n + 2324 \over {162 \pi}}
\Biggl({\lambda Z^2 \over m} \Biggr)  \qquad \qquad \qquad 
\qquad \qquad \qquad \qquad 
\nonumber \\
+ (0.0056289546468~n^3 + 0.28932672886~n^2 +
4.0404241235~n + 16.204286853){\Biggl({\lambda Z^2 \over m} 
\Biggr)^2}  
\nonumber \\
- (0.001493126~n^4 + 0.09961447~n^3 + 2.152320~n^2 
+ 18.330704~n + 52.830284)
{\Biggl({\lambda Z^2 \over m} \Biggr)^3} \Biggr].  
\label{eq:9}
\end{eqnarray} 
The perturbative expansion for $\lambda$ emerges directly from the 
normalizing condition $\lambda = m Z_4 Z^{-2} g_4$ and the known series 
for $Z_4$ \cite{AS}:
\begin{eqnarray}
Z_4 = 1  +  { n + 8 \over {2 \pi}} g_4  +  
{3~n^2 + 38~n + 148 \over {12 \pi^2}} g_4^2  
\qquad \qquad \qquad \qquad \qquad \qquad \qquad \qquad \qquad
\nonumber \\
+ (0.0040314418~n^3 + 0.0679416657~n^2 
+ 0.466356233~n + 1.240338484) g_4^3.
\label{eq:10}
\end{eqnarray} 
Combining these expressions, we obtain
\begin{eqnarray}
g_6 = {9 \over \pi} g_4^3 \Biggl[ {n + 26 \over {27}} -  
{17~n + 226 \over {81 \pi}} g_4 + (0.000999164~n^2 
+ 0.14768927~n + 1.24127452) g_4^2 
\nonumber \\
-~(- 0.00000949~n^3 + 0.00783129~n^2 + 0.34565683~n 
+ 2.14825455) g_4^3 \Biggr]. \qquad \qquad 
\label{eq:11}
\end{eqnarray} 

In the case of $g_8$, the one-, two-, and three-loop contributions are
given by one, five, and 36 Feynman graphs, respectively. Corresponding 
"bare" and renormalized perturbative expansions are found to be: 
\begin{eqnarray}
g_8 = -{81 \over {2 \pi}}{\Biggl({\lambda Z^2 \over m} \Biggr)^4}
\Biggl[ {n + 80 \over {81}}  -  {405~n^2 + 35626~n + 342320 \over 
{13122 \pi}}\Biggl({\lambda Z^2 \over m} \Biggr)  \qquad \qquad \qquad 
\qquad \qquad 
\nonumber \\
+~(0.0046907955~n^3 + 0.463650683~n^2 + 8.86811653~n 
+ 45.4769028){\Biggl({\lambda Z^2 \over m} \Biggr)^2} \Biggr]. \qquad  
\label{eq:12}
\end{eqnarray}
\begin{eqnarray}
g_8 = -{81 \over {2 \pi}} g_4^4 \Biggl[ {n + 80 \over {81}} -  
{81~n^2 + 7114~n + 134960 \over {13122 \pi}} g_4 
\qquad \qquad \qquad \qquad
\nonumber \\
+ (0.00943497~n^2 + 0.60941312~n + 7.15615323) g_4^2 \Biggr]. 
\qquad 
\label{eq:13}
\end{eqnarray} 

In Sec.III, the series Eqs.(\ref{eq:11}), (\ref{eq:13}) will 
be used for estimation of the universal numbers $g_6^*$ and $g_8^*$.
  
\section{Resummation and numerical estimates}
\label{sec:3}

Being a field-theoretical perturbative expansions the series of equations 
(\ref{eq:11}), (\ref{eq:13}) have factorially growing coefficients, 
i. e., they are divergent (asymptotic). Hence, direct substitution of 
the fixed point value $g_4^*$ into them would not lead to satisfactory 
results. To get reasonable numerical estimates for $g_6^*$ and $g_8^*$, 
some procedure making these expansions convergent should be applied. 
As is well known, the Borel-Leroy transformation 
\begin{equation}
f(x) =  \sum_{i = 0}^{\infty} c_i x^i = 
\int\limits_0^{\infty} t^b e^{-t} F(xt) dt , \qquad
F(y) = \sum_{i = 0}^{\infty} {c_i \over (i+b)!} y^i ,
\label{eq:14}
\end{equation}
diminishing the coefficients by the factor $(i+b)!$, can play a role 
of such a procedure. Since the RG series considered turns out to be 
alternating the analytical continuation of the Borel transform may 
be then performed by using Pad\'e approximants. 

Let us discuss first the estimation of the sextic coupling constant $g_6^*$. 
With the four-loop expansion (\ref{eq:11}) in hand, we can construct, 
in principle, three different Pad\'e approximants: [2/1], [1/2], 
and [0/3]. To obtain proper approximation schemes, however, only diagonal 
[L/L] and near-diagonal Pad\'e approximants should be employed 
\cite{BGM}. That is why, further, when estimating $g_6^*$, we limit 
ourselves with approximants [2/1] and [1/2]. Moreover, the diagonal 
Pad\'e approximant [1/1] will be also dealt with, although 
this corresponds, in fact, to the usage of the lower-order, three-loop 
RG approximation. 

The algorithm of estimating $g_6^*$ we use here is as follows. 
Since the Taylor expansion for the free energy contains as 
coefficients the ratios $R_{2k} = g_{2k}/g_4^{k-1}$ rather than 
the renormalized coupling constants themselves, we work with the RG 
series for $R_6$. It is resummed in three different ways based on the 
Borel-Leroy transformation and the Pad\'e approximants just 
mentioned. The Borel-Leroy integral is evaluated as a function of 
the parameter $b$ under $g_4 = g_4^*$. For the fixed point coordinate 
$g_4^*$, the values given by the resummed six-loop RG expansion for the 
$\beta$-function are adopted \cite{BNM,S98},   
which are believed to be the most accurate estimates available today. 
The optimal value of $b$ providing the fastest convergence of the 
iteration scheme is then determined. It is deduced from the condition 
that the Pad\'e approximants employed should give, for $b = b_{opt}$, 
the values of $R_6^*$ which are as close as possible to each other. 
Finally, the average over three estimates for $R_6^*$ is found and 
claimed to be a numerical value of this universal ratio. 
 
To obtain an idea about how such a procedure works, let us use 
Table I, where the results of corresponding calculations for $n = 1$, 
3, and 10 are presented. It is seen that for $n = 1$ and 
3, ~$b_{opt}$, providing a coincidence of the estimates given by 
all three working Pad\'e approximants, is equal to 1.24. For 
$n = 10$, $b_{opt}$, fixed by the approximants [1/1] and [2/1], is equal to 
1, whereas the third approximant ([1/2]) at $b = 1$ is spoiled by a positive 
axis pole. Nevertheless, the numerical estimate given by this approximant 
under the nearest "safe" (integer) value of $b$ ($b = 2$) turns out to be 
very close to that predicted by the pole free approximants for $b_{opt}$. 
Moreover, as is seen from Table I, with increasing $n$ numerical estimates 
for $g_6^*$ become less dependent on $b$, i. e., their sensitivity 
to the type of resummation decreases. This is not surprising. The point 
is that the RG expansion (\ref{eq:11}) becomes less divergent when $n$ 
grows up. To make this property obvious, let us replace $g_4$ in 
Eq.(\ref{eq:11}) by the effective coupling constant  
\begin{equation}
g = {{n + 8} \over {2 \pi}} g_4,
\label{eq:15}
\end{equation}    
that is known to be only weakly dependent on $n$: it varies from 1.415 
to 1 when $n$ goes from 1 to infinity \cite{AS,S98}. Then we obtain         
\begin{eqnarray}
g_6 = {{8 \pi^2 (n + 26)} \over {3 (n + 8)^3}}~g^3~\Biggl[1 - 
{2(17 n + 226) \over {3(n + 8)(n + 26)}}~g
+ {{1.065025 n^2 + 157.42454 n + 1323.09596} \over{(n + 8)^2 (n + 26)}}~g^2
\nonumber \\
- {{-0.0638 n^3 + 52.4510 n^2 + 2314.9897 n + 14387.6460} 
\over{(n + 8)^3 (n + 26)}}~g^3 \Biggr]. \ \   
\label{eq:16}
\end{eqnarray}
One can see now that all the terms in the RG expansion for $g_6$ 
(in square brackets), apart from the first one, decrease monotonically  
when $n \to \infty$. This implies that the larger $n$ is the smaller the 
contribution of the higher-order terms and, correspondingly,
the better the approximating properties of this series.

This conclusion is definitely confirmed by Table II. It contains numerical 
estimates for $g_6^*$ resulting from the four-loop RG expansion resummed by 
the Pad\'e-Borel-Leroy technique described above (column 3) and  
their analogs given by the Pad\'e-Borel resummed three-loop 
RG series \cite{S98} (column 4). As is seen, with increasing $n$ the 
difference between the four-loop and three-loop estimates rapidly 
diminishes. Being small (0.9 \%) even for $n = 1$, it becomes negligible 
at $n = 10$ and practically disappears for $n \ge 14$.   

How close to the exact values of $g_6^*$ may the numbers in column 3 be?  
To clear up this point, let us compare our four-loop estimate for $R_6^*$ 
at $n = 1$ with those obtained recently by an analysis of the five--loop 
scaling equation of state for the 3D Ising model \cite{GZ,GZ98}. Guida 
and Zinn-Justin obtained $R_6^* = 1.644$ and, taking into 
account some additional information, $R_6^* = 1.643$, while our estimate is 
$R_6^* = 1.648$. Keeping in mind that the exact value of $R_6^*$ should lie 
between the four-loop and five-loop estimates (the RG series is 
alternating), our estimate can differ from the exact number by no more 
than 0.3 \%. Since for $n > 1$ the RG expansion (\ref{eq:11}) was argued 
to provide the better numerical estimates than in the Ising case, this 
value (0.3 \%) may be referred to as an upper bound for the deviation of 
the numbers in column 3 of Table II from their exact counterparts. 
     
It is interesting to compare our estimates for $g_6^*$ with those obtained 
by other methods. Since 1994, the universal values of the sextic coupling 
constant for the 3D $O(n)$-symmetric model were estimated by solving 
the exact RG equations \cite{TW}, by lattice calculations \cite{R}, and by 
a constrained analysis of the $\epsilon$-expansion \cite{PV}; corresponding 
results are collected in columns 5, 6, and 7 of Table II, respectively. 
As is seen, they are, in general, in accord with ours.

A less optimistic situation takes place in the case of the octic 
coupling constant $g_8$. The RG expansion Eq.(\ref{eq:13}) is 
shorter than Eq.(\ref{eq:11}), and stronger diverges. Moreover, the second 
term in this series, along with the first one, remains finite under 
$n \to \infty$. It becomes obvious if one replaces $g_4$ by $g$:  
\begin{eqnarray}
g_8 = -{{8 \pi^3 (n + 80)} \over{(n + 8)^4}}~g^4 \Biggl[1 -  
{{81~n^2 + 7114~n + 134960} \over {81(n + 80)(n + 8)}}~g 
\qquad \qquad \qquad \qquad
\nonumber \\
+ {{30.1707~n^2 + 1948.7519~n + 22883.6021} \over {(n + 80)(n + 8)^2}}~g^2 
\Biggr]. \qquad \qquad 
\label{eq:17}
\end{eqnarray}
In addition, the RG series for $g_8$ has an unusual feature: when 
$n \to \infty$, the first and second terms tend to compensate each other, 
making their mutual contribution small and increasing the role of the 
higher-order terms. That is why numerical estimates resulting from the 
expansion (\ref{eq:17}) are expected to be substantially cruder than those 
given by the series (\ref{eq:16}) both for small and large values of $n$.

In order to estimate $g_8^*(n)$, we resum the RG expansion for $g_8$ by the 
Pad\'e-Borel-Leroy technique using the diagonal Pad\'e 
approximant [1/1]. Other Pad$\acute e$ approximants, [0/2] and [0/1], are 
ignored, since they turn out to lead to quite unsatisfactory numerical 
results. Dealing with a single Pad$\acute e$ approximant,  
in some condition we need to fix the optimal value of the shift 
parameter $b$. For the three-dimensional Ising model the estimate 
$g_8^* = 0.825$ was recently found \cite{GZ}. This number has been 
extracted from the five-loop RG expansion, and may be considered the most 
accurate known up to the present. It is natural therefore to tune, 
by proper choice of $b$, a numerical value of $g_8^*(1)$ given by the 
resummed three-loop RG series with the best estimate available. Such a 
procedure leads to $b_{opt} = 40$, and this number is adopted as optimal 
in the course of evaluation of $g_8^*$ for arbitrary $n$. 

The results of our calculations are collected in Table III,  
where the estimates for $g_8^*(n)$, obtained by a constrained analysis of the 
$\epsilon$-expansion \cite{PV}, by approximate solution of the exact RG 
equations \cite{TW}, and given by the $1/n$-expansion technique, are also 
presented for comparison. As seen, 
for $n \ge 8$ the numbers originating from two field-theoretical approaches 
-- $g$-expansion in three dimensions and $\epsilon$-expansion -- agree quite 
well. However, for smaller $n$, especially for $n = 2$, differences between 
them turn out to be rather large. This is not surprising since overly short 
perturbative expansions for $g_8$ are available both in 3 and $4 - \epsilon$ 
dimensions and they demonstrate a strong divergence preventing accurate 
numerical estimates from being obtained. At the same time, our three-loop 
RG estimates are believed to be closer to the true critical values of $g_8$ 
than those given by the $\epsilon$-expansion, because in three dimensions 
we have longer perturbative series. A fair agreement between our results 
and the numbers emerging from the exact RG equations (see Table III) may be 
considered as an argument in favor of this belief.  

\section{Conclusion}    
\label{sec:4}
    
To summarize, we have calculated the RG expansions for renormalized coupling  
constants $g_6$ and $g_8$ of the 3D $n$-vector model in four-loop 
and three-loop orders, respectively. Resummation of the RG series by the 
Pad\'e-Borel-Leroy method has enabled us to obtain numerical estimates 
for the universal critical values of these quantities for arbitrary $n$. 
Having analyzed the sensitivity of the RG estimates for $g_6^*$ to the type 
of resummation procedure and a character of their dependence on the order of 
the RG approximation, an apparent accuracy of these numbers has been argued 
to be no worse than 0.3$\%$. Numerical estimates for $g_8^*$ turned out to 
be much less accurate because of a smaller length and stronger 
divergence of the RG expansion obtained. They were found to be consistent,
in general, with the values of $g_8^*$ deduced from the exact RG equations 
and, for $n \ge 8$, with those given by a constrained analysis of 
corresponding $\epsilon$-expansion.

\acknowledgments 
 
This work was supported by the Ministry of General and Professional 
Education of Russian Federation under Grant No. 97--14.2--16. One of 
the authors (A. I. S.) gratefully acknowledges the support of 
the International Science Foundation via Grant p98--537.

\newpage
\widetext
\begin{table}
\caption{The values of $g_6^*$ for $n = 1$, 3, and 10 
obtained by means of the Pad\'e-Borel-Leroy 
technique for various $b$ within three-loop (approximant $[1/1]$) 
and four-loop (approximants $[1/2]$ and $[2/1]$) RG approximations. 
The estimates for several values of 
$b$ in the middle lines are absent because corresponding Pad\'e
approximant turns out to be spoiled by a positive axis pole.}
\begin{tabular}{ccccccccc}
$b$ & 0 & 1 & 1.24 & 2 & 3 & 4 & 5 & 7 \\
\tableline
n = 1 \\
\tableline
$[1/1]$ & 1.576 & 1.604 & 1.6089 & 1.621 & 1.633 & 1.641 & 1.648 & 1.656 \\
\tableline
$[1/2]$ & - & - & 1.6084 & 1.600 & 1.595 & 1.592 & 1.590 & 1.587 \\
\tableline
$[2/1]$ & 1.639 & 1.613 & 1.6084 & 1.596 & 1.583 & 1.573 & 1.566 & 1.555 \\
\tableline
n = 3 \\
\tableline
$[1/1]$ & 0.937 & 0.949 & 0.95133 & 0.957 & 0.962 & 0.966 & 0.969 & 0.973 \\
\tableline
$[1/2]$ & - & - & 0.95133 & 0.948 & 0.946 & 0.944 & 0.944 & 0.942 \\
\tableline
$[2/1]$ & 0.964 & 0.953 & 0.95133 & 0.946 & 0.941 & 0.937 & 0.934 & 0.930 \\
\tableline
n = 10 \\
\tableline
$[1/1]$ & 0.2338 & 0.23515 &  & 0.2360 & 0.2366 & 0.2370 & 0.2373 & 0.2377 \\
\tableline
$[1/2]$ & - & - &  & 0.2348 & 0.2346 & 0.2345 & 0.2344 & 0.2342 \\
\tableline
$[2/1]$ & 0.2359 & 0.23515 &  & 0.2346 & 0.2342 & 0.2339 & 0.2337 & 0.2334 \\
\end{tabular}
\end{table}

\begin{table}
\caption{Our estimates of universal critical values of the renormalized 
sextic coupling constant for the 3D $n$-vector model (column 3). 
The fixed point coordinates $g^*$ are taken from Ref.[3] 
($1 \le n \le 3$) and Ref.[7] ($4 \le n \le 40$).
The $g_6^*$ estimates extracted earlier from Pad\'e-Borel resummed  
three-loop RG expansion (column 4), from the exact RG equations 
(column 5), obtained by the lattice calculations (column 6), 
and resulting from a constrained analysis of the  
$\epsilon$-expansions (column 7) are presented for comparison. 
Column 8 contains the values of $g_6^*$ given by 
the $1/n$-expansion technique.} 
\begin{tabular}{c||c||c|c|c|c|c|c}
\hline
$n$ & $g^*$ & $g_6^*$ & $g_6^* \cite{S98}$ & $g_6^* \cite{TW}$ & 
$g_6^* \cite{R}$ & $g_6^* \cite{PV}$ & $g_6^*~(1/n)$ \\
\hline
  & 2 & 3 & 4 & 5 & 6 & 7 & 8 \\
\hline
1 & 1.415 & 1.608 & 1.622 & 1.52 & 1.92(24) & 1.609(9) & \\
\hline
2 & 1.406 & 1.228 & 1.236 & 1.14 & 1.27(25) & 1.21(7) & \\
\hline
3 & 1.392 & 0.951 & 0.956 & 0.88 & 0.93(20) & 0.931(46) & \\
\hline
4 & 1.3745 & 0.747 & 0.751 & 0.68 & 0.62(15) & 0.725(29) & 1.6449 \\
\hline
5 & 1.3565 & 0.596 & 0.599 &  &   &  & 1.0528 \\
\hline
6 & 1.3385 & 0.483 & 0.485 &  &   &  & 0.7311 \\
\hline
7 & 1.321 & 0.396 & 0.398 &   &   &  & 0.5371 \\
\hline
8 & 1.3045 & 0.329 & 0.331 &   &   & 0.319(4) & 0.4112 \\
\hline
9 & 1.289 & 0.277 & 0.278 &   &   &  & 0.3249 \\
\hline
10 & 1.2745 & 0.235 & 0.236 &   &   &  & 0.2632 \\
\hline
12 & 1.2487 & 0.174 & 0.175 &   &   &  & 0.1828 \\
\hline
14 & 1.2266 & 0.134 & 0.134 &   &   &  & 0.1343 \\
\hline
16 & 1.2077 & 0.105 & 0.105 &   &   & 0.1032(4) & 0.1028 \\
\hline
18 & 1.1914 & 0.0845 & 0.0847 &   &   &  & 0.0812 \\
\hline
20 & 1.1773 & 0.0693 & 0.0694 &   &   &  & 0.0658 \\
\hline
24 & 1.1542 & 0.0487 & 0.0488 &   &   &  & 0.0457 \\
\hline
28 & 1.1361 & 0.0360 & 0.0361 &   &   &  & 0.0336 \\
\hline
32 & 1.1218 & 0.0276 & 0.0276 &   &   & 0.0275(1) & 0.0257 \\
\hline
36 & 1.1099 & 0.0218 & 0.0218 &   &   &  & 0.0203 \\
\hline
40 & 1.1003 & 0.0176 & 0.0176 &   &   &  & 0.0164 \\
\hline
\end{tabular}
\end{table}

\begin{table}
\caption{Three-loop RG estimates of universal critical values of 
the renormalized octic coupling constant $g_8$ (column 2).  
The $g_8^*$ estimates resulting from a constrained analysis of the  
$\epsilon$-expansion (column 3), from the exact RG 
equations (column 4) and given by the $1/n$-expansion 
technique (column 5) are presented for comparison.}
\begin{tabular}{c|c|c|c|c}
\hline
$n$ & $g_8^*$ & $g_8^*~\cite{PV}$ & $g_8^*~\cite{TW}$ & $g_8^*~(1/n)$ \\
\hline
  & 2 & 3 & 4 & 5 \\
\hline
1 & 0.825 & 0.82(9) & 0.721 & \\
\hline
2 & 0.388 & 0.83(31) & 0.343 & \\
\hline
3 & 0.168 & 0.36(17) & 0.145 & \\
\hline
4 & 0.057 & 0.15(13) & 0.042 & -2.151 \\
\hline
6 & -0.021 &   &  & -0.834 \\
\hline
8 & -0.034 & -0.03(2) &  & -0.0388 \\
\hline
16 & -0.014 & -0.015(2) &  & -0.0456 \\
\hline
32 & -0.0023 & -0.0023(1) &  & -0.00395 \\
\hline
48 & -0.00062 & -0.00061(2) &  & -0.00087 \\
\hline
64 & -0.00023 &  &  & -0.00029 \\
\hline
100 & -0.000046 & -0.000044(2) & -0.000049 & -0.000052 \\
\hline
\end{tabular}
\end{table}
\end{document}